\documentclass[twoside,reqno]{bjp}
\usepackage{epsfig,cite}
\usepackage{graphics}
\usepackage{amssymb,amsmath}
\usepackage{times}
\begin{document}

\sloppy \raggedbottom

 \setcounter{page}{1}

\title{Foundations of the proxy-SU(3) symmetry in heavy nuclei}

\runningheads{Foundations of the proxy-SU(3) symmetry in heavy nuclei}{I.E. Assimakis,
D. Bonatsos, N. Minkov, A. Martinou, R.B. Cakirli, et al.}

\begin{start}

\author{I.E. Assimakis}{1},
\coauthor{D. Bonatsos}{1},
\coauthor{N. Minkov}{2},
\coauthor{A. Martinou}{1},
\coauthor{R.B. Cakirli}{3},
\coauthor{R.F. Casten}{4,5},
\coauthor{K. Blaum}{6}

\address{Institute of Nuclear and Particle Physics, National Centre for Scientific Research ``Demokritos'', GR-15310 Aghia Paraskevi, Attiki, Greece}{1}

\address{Institute of Nuclear Research and Nuclear Energy, Bulgarian Academy of Sciences, 72 Tzarigrad Road, 1784 Sofia, Bulgaria}{2}

\address{Department of Physics, University of Istanbul, 34134 Istanbul, Turkey}{3} 

\address{Wright Laboratory, Yale University, New Haven, Connecticut 06520, USA}{4}

\address{Facility for Rare Isotope Beams, 640 South Shaw Lane, Michigan State University, East Lansing, MI 48824 USA}{5}

\address{Max-Planck-Institut f\"{u}r Kernphysik, Saupfercheckweg 1, D-69117 Heidelberg, Germany}{6}

\received{31 October 2017}

\begin{Abstract}
We show that within the proxy-SU(3) scheme the wave functions of the normal parity orbitals in a given nuclear shell are affected very little as a result of the replacement of the abnormal parity orbitals by their 0[110] proxy-SU(3) counterparts.

\end{Abstract}

\end{start}

\section{Introduction}

An approximate SU(3) symmetry, called the proxy-SU(3) symmetry \cite{proxy1}, appears in heavy deformed even-even nuclei, by omitting the intruder Nilsson orbital \cite{Nilsson1,Nilsson2} of highest total
angular momentum and replacing the rest of the intruder orbitals by the orbitals which have escaped to the next lower major shell. 
The approximation is based on the fact that there is a one-to-one correspondence between the orbitals of the two sets, based on pairs 
of orbitals having identical quantum numbers of orbital angular momentum, spin, and total angular momentum \cite{proxy1,Karampagia}. These are called 0[110] pairs \cite{Cakirli}, where the numbers stand for the change in the $K[N n_z \Lambda]$ aymptotic quantum numbers of the Nilsson model \cite{Nilsson1,Nilsson2}, 
 where $N$ is the total number of oscillator quanta, 
$n_z$ is the number of the oscillator quanta along the $z$-axis, $\Lambda$ is the 
$z$-projection of the orbital angular momentum, and $K$ (for even nuclei) 
is the $z$-projection of the total angular momentum.
The accuracy of the approximation 
is tested through calculations in the framework of the Nilsson model \cite{Nilsson1,Nilsson2} in the asymptotic limit of large deformations, focusing attention on the changes 
in selection rules and in avoided crossings caused by the opposite parity of the proxies with respect to the substituted  orbitals \cite{proxy1}. 

In Section 2 of the present work, the proxy-SU(3) wave functions for the normal parity orbitals are compared to the usual Nilsson orbitals, showing that the overlap between them remains large. 

\section{Wave functions in the proxy-SU(3) Nilsson calculation}\label{wf}

In Ref.  \cite{proxy1} the proxy-SU(3) scheme has been justified in terms of two different Nilsson calculations using the asymptotic Nilsson wave functions \cite{Nilsson1,Nilsson2}. 
In one of the calculations the usual orbitals occurring in the 50-82 proton shell 
have been used, while in the other, the orbitals occurring in the full sdg shell 
of proxy-SU(3) have been implemented (see Table V of \cite{proxy1}).
The results of the full diagonalization have been shown in Fig. 2 of Ref. \cite{proxy1}, in which it is seen that the Nilsson diagrams resulting in the two cases are very similar, due to the small size of the additional matrix elements appearing in the latter calculation \cite{proxy1}.

A reasonable question regards the structure of the wave functions in the two calculations. The following comments apply:

1) The 50-82 calculation involves 10 Nilsson orbitals of normal parity (positive in this case) and 6 orbitals of opposite parity, while the sdg proxy calculation involves the same 10 positive parity orbitals as before, plus five more orbitals of positive parity, as seen in Table V of Ref. \cite{proxy1}. 

2) One can use as basis for the comparison the Nilsson orbitals of the full sdg shell ($N=4$). It should be noticed that the components of the 50-82 wave functions 
involving the negative parity orbitals will not make any contribution in this case,
since the involved orbitals from the next higher shell ($N=5$) are orthogonal to the sdg orbitals. 

3) As a result, when calculating the overlaps of the 50-82 and sdg wave functions
we are going to get nonvanishing contributions only from the 10 positive parity orbitals appearing in both shells. 

\begin{table*}

\caption{Overlaps of wave functions for the 50-82 proton shell with the sdg proxy-SU(3) sdg shell. See section \ref{wf} for further discussion. 
}

\bigskip

\begin{tabular}{ r r r r r r r r  }
                  
$\epsilon$ & 0.0  & 0.1 & 0.2 & 0.3 & 0.4 & 0.5 & 0.6 \\

\hline

 1/2[400] &  0.999  & 0.994 & 0.853 & 0.997 & 1.000 & 1.000 & 1.000 \\
 1/2[411] &  0.981  & 0.971 & 0.932 & 0.670 & 0.965 & 0.992 & 0.997 \\
 3/2[402] &  0.977  & 0.935 & 0.846 & 0.984 & 0.996 & 0.998 & 0.999 \\
 1/2[420] &  0.987  & 0.996 & 0.990 & 0.969 & 0.895 & 0.722 & 0.892 \\
 3/2[411] &  0.997  & 0.988 & 0.962 & 0.879 & 0.732 & 0.892 & 0.949 \\
 5/2[402] &  0.992  & 0.970 & 0.888 & 0.782 & 0.933 & 0.972 & 0.985 \\
 1/2[431] &  0.998  & 0.980 & 0.971 & 0.957 & 0.935 & 0.900 & 0.838 \\
 3/2[422] &  0.967  & 0.953 & 0.933 & 0.903 & 0.856 & 0.782 & 0.727 \\
 5/2[413] &  0.950  & 0.929 & 0.896 & 0.842 & 0.755 & 0.767 & 0.856 \\
 7/2[404] &  0.959  & 0.931 & 0.873 & 0.759 & 0.805 & 0.898 & 0.943 \\

 \hline
\end{tabular}
\end{table*}

In Table 1 the overlaps of the wave functions corresponding to the 50-82 shell 
and to the full sdg proxy-SU(3) shell are shown for 7 different values of the deformation 
parameter $\epsilon$. The order of the orbitals is kept the same as in Table V of Ref. \cite{proxy1}. 

The notation in Table 1 has the following meaning. Because in both the 50-82 Nilsson calculation and the sdg proxy calculation the non-diagonal matrix elements 
of the $l\cdot s$ and $l^2$ terms are small, one can keep track of the evolution 
of each Nilsson orbital with increasing values of $\epsilon$. For a given value 
of $\epsilon$, the wave function of a given orbital (the one carrying the name 5/2[402], for example),
will contain in both calculations a main component corresponding to the orbital 
of which the name it bears, plus small components on the other members of the orthogonal basis. We can then easily calculate the overlap of the two wave functions. 

Table 1 indicates that the replacement of the negative parity orbitals by their 
0[110] proxies does have some influence on the positive parity orbitals, but 
the positive parity orbitals keep their character to a large extent. This is in accordance with the small size of the matrix elements added because of the proxy replacement, as seen in Tables III-V of Ref. \cite{proxy1}.

In different words, it appears that the main role of the spin-orbit interaction 
is to modify the shell borders and change the magic numbers from the 3D harmonic oscillator ones to the nuclear magic numbers. Once this is achieved, 
the replacement of the opposite parity orbitals by their 0[110] proxies
is a good approximation, at least for even-even nuclei, in which the parity 
of the orbitals does not play a significant role for a number of properties. It is a good approximation because all angular momentum projections remain intact. Except in cases 
in which parity plays a major role (odd nuclei, for example), the proxy orbitals will play the same role as the original orbitals which they replaced. 

It should be reminded \cite{proxy1} that the 50-82 shell is {\sl the worst} possible example for examining the changes inflicted by the 0[110] proxy substitution. The changes get smaller in higher shells, making the proxy approximation even better. It should also be taken into account that the strength of the spin-orbit interaction is expected from relativistic mean field calculations to be significantly reduced \cite{Lal} away from the stability line, thus making the proxy predictions even more accurate.  

In corroboration of the above discussion, we give in Tables II-V the matrix elements 
of the spin-orbit interaction, the $l^2$ angular momentum term and the total Hamiltonian 
for both the normal Nilsson calculation and the proxy pfh calculation, corresponding to the 82-126 shell (see Ref. \cite{proxy1} for further details). These tables have been deposited as supplemental material for Ref. 
\cite{proxy1}, as described in Ref. [33] of \cite{proxy1}. Indeed one sees that the modified matrix elements are very few and their effect 
on the Hamiltonian is very small.

It should be emphasized that the Nilsson calculation in the proxy-SU(3) case 
is not aiming at proving that we can use the Nilsson Hamiltonian for any proxy-SU(3) calculations. We do not try to modify the well known Nilsson diagrams. We are just 
using this Nilsson calculation as an argument that SU(3) can be applied, 
within the proxy-SU(3) framework, to heavy deformed nuclei. The relevant algebraic 
Hamiltonian will be of the form $H=Q\cdot Q + h$, where $h$ is a higher order term, 
either $\Omega$ or $\Lambda$ \cite{PVI}. Then one has a full set of operators and wave functions, thus one can make calculations of physical quantities and obtain parameter-free predictions, which can be compared to the data.

\section{Discussion}

The fact that the structure of the wave functions has to be similar 
when one tries to establish agreement between two models/approaches/approximations
is easily understood. We have proved that the proxy-SU(3) approximation affects little 
the wave functions of the normal parity orbitals in a given shell.   

In nuclei it seems that the major role of the spin-orbit interaction is to 
define the borders (shell gaps). Within each region, most of the work is then done by the Pauli principle (and by parity in odd nuclei). The Pauli principle is always present in nuclei, doing its job irrespectively of the form of any Hamiltonian, 
bosonic or fermionic. The explanation of the prolate over oblate dominance as an application of the proxy-SU(3) scheme is a 
spectacular demonstration \cite{proxy2,EPJA} of this fact.  
  
An appropriate test ground for the proxy-SU(3) scheme are the M1 transition probabilities and the magnetic moments, since they involve single-particle operators. This test ground has been used, for example, in \cite{pseudo1,Yoshida}.

\section*{Acknowledgements}

Work partly supported by the Bulgarian National Science Fund (BNSF) under Contract No. DFNI-E02/6, by the US DOE under Grant No. DE-FG02- 91ER-40609, and by the MSU-FRIB laboratory, by the Max Planck Partner group, TUBA-GEBIP, and by the Istanbul University Scientific Research Project No. 54135.


\begin{figure}[b]
\centering{\epsfig{file=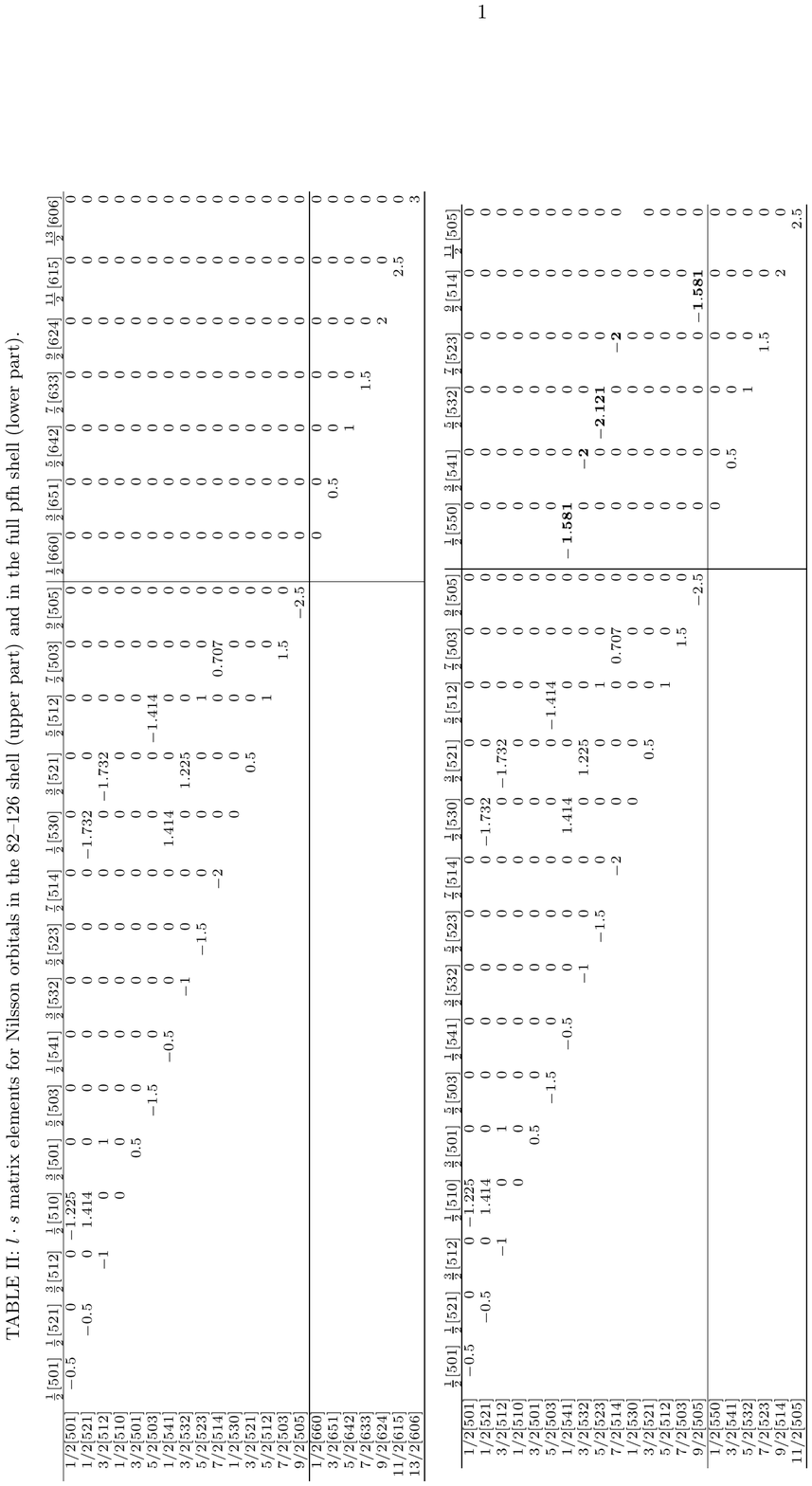,width=140mm}}

\end{figure}


\begin{figure}[b]
\centering{\epsfig{file=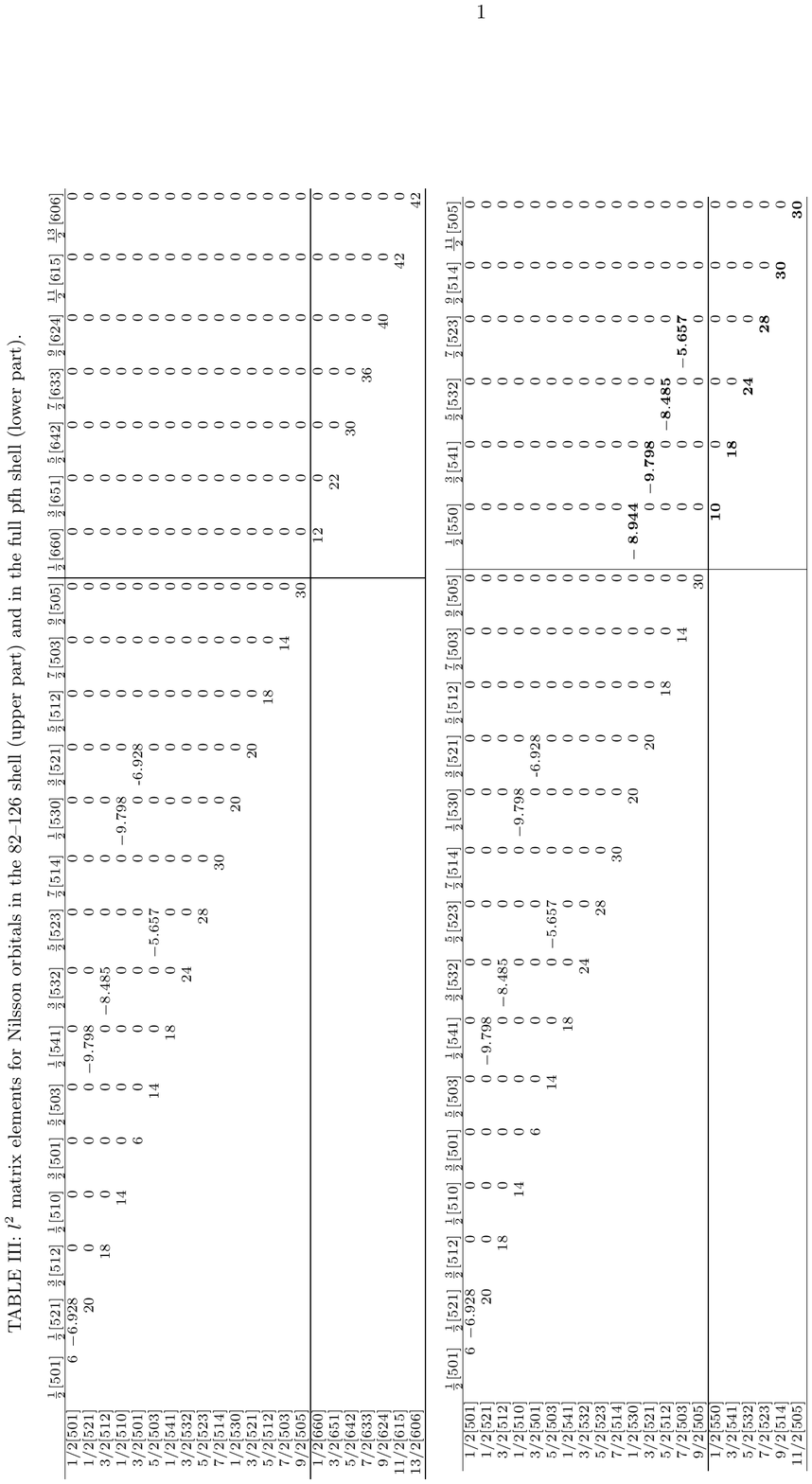,width=140mm}}

\end{figure}


\begin{figure}[b]
\centering{\epsfig{file=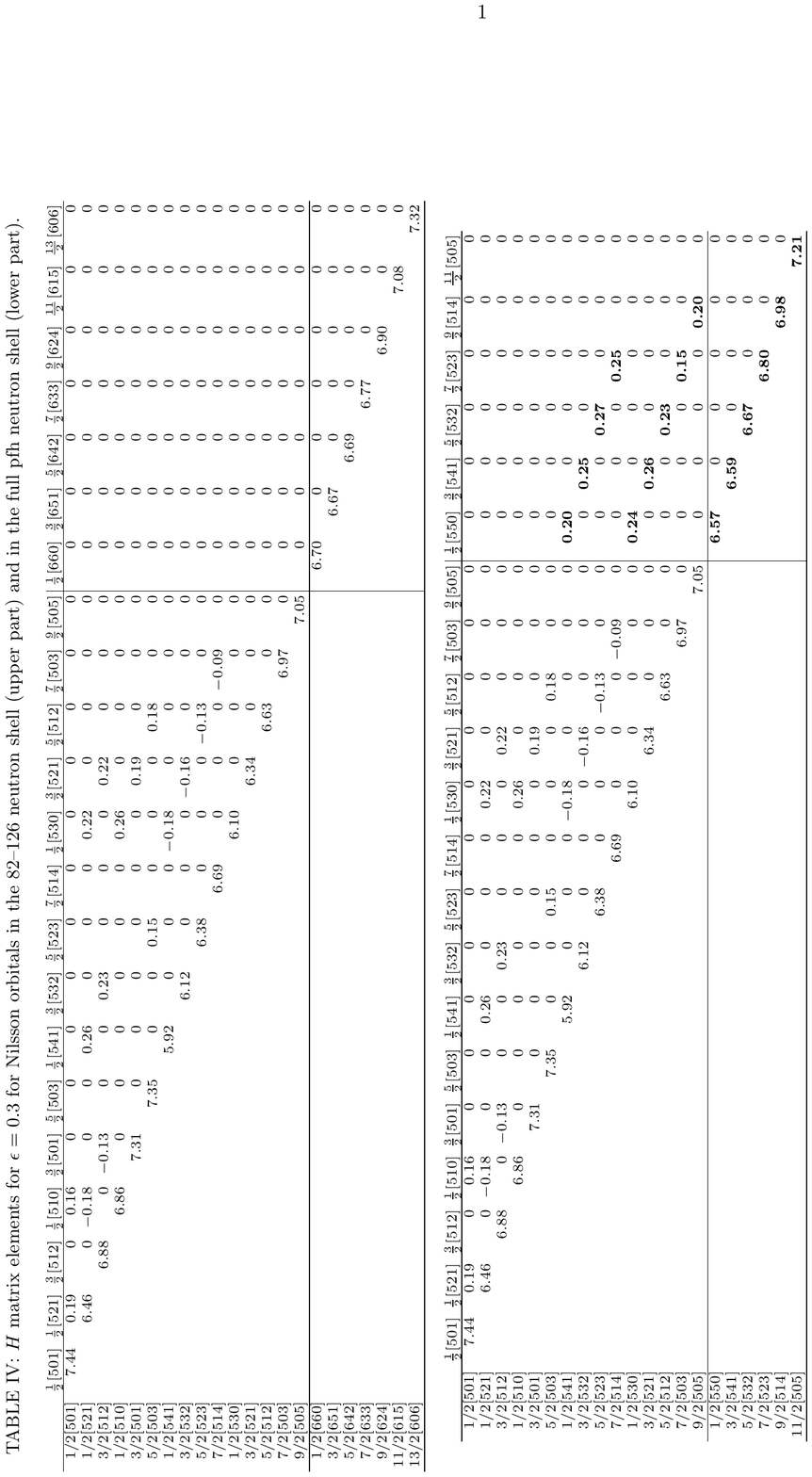,width=140mm}}

\end{figure}


\begin{figure}[b]
\centering{\epsfig{file=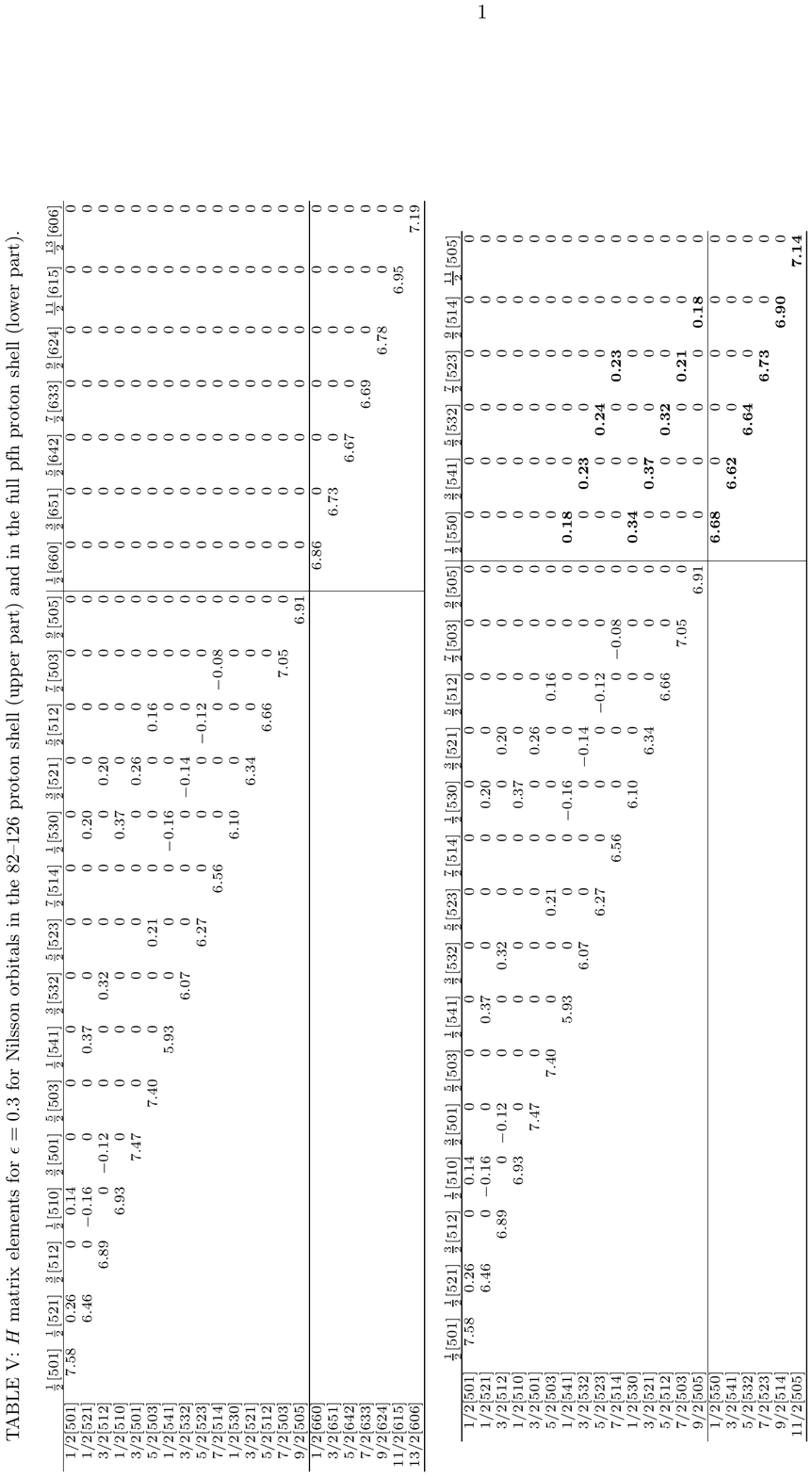,width=140mm}}

\end{figure}


\begin{thebibliography}{99}

\bibitem{proxy1}
D. Bonatsos, I. E. , N. Minkov, A. Martinou, R. B. Cakirli, R. F. Casten, and K. Blaum, Proxy SU(3) symmetry in heavy deformed nuclei, Phys. Rev. C \textbf{95}, 064325 (2017).

\bibitem{Nilsson1} 
S. G. Nilsson, Binding states of individual nucleons in strongly deformed nuclei, 
Mat. Fys. Medd. K. Dan. Vidensk. Selsk. \textbf{29}, no. 16 (1955).

\bibitem{Nilsson2} 
S. G. Nilsson and I. Ragnarsson, Shapes and Shells in Nuclear Structure (Cambridge University Press,
Cambridge, 1995). 

\bibitem{Karampagia}   
D. Bonatsos, S. Karampagia, R. B. Cakirli, R. F. Casten, K. Blaum, and L. Amon Susam, 
Emergent collectivity in nuclei and enhanced proton-neutron interactions, Phys. Rev. C \textbf{88}, 054309 (2013).

\bibitem{Cakirli}
R. B. Cakirli, K. Blaum, and R. F. Casten, Indication of a mini-valence Wigner-like energy in heavy nuclei, 
Phys. Rev. C \textbf{82}, 061304(R) (2010). 

\bibitem{Lal}
D. Vretenar, A. V. Afanasjev, G. A. Lalazissis, and P. Ring, Relativistic Hartree-Bogoliubov 
theory: Static and dynamic aspects of exotic nuclear structure, Phys. Rep. \textbf{409},101 (2005). 

\bibitem{PVI}
G. Vanden Berghe, H. E. De Meyer, and P. Van Isacker, Symmetry-conserving higher-order interaction terms in the interacting boson model, Phys. Rev. C \textbf{32}, 1049 (1985).

\bibitem{proxy2}
D. Bonatsos, I. E. Assimakis, N. Minkov, A. Martinou, S. Sarantopoulou, R. B. Cakirli, R. F. , and K. Blaum, Analytic predictions for nuclear shapes, prolate dominance and the prolate-oblate shape transition in the proxy-SU(3) model, Phys. Rev. C \textbf{95}, 064326 (2017).

\bibitem{EPJA} 
D. Bonatsos, Prolate over oblate dominance in deformed nuclei as a consequence of the SU(3) symmetry and the Pauli principle, Eur. Phys. J. A \textbf{53}, 148 (2017). 

\bibitem{pseudo1}
R. D. Ratna Raju, J. P. Draayer, and K. T. Hecht, Search for a coupling scheme in heavy deformed nuclei: The pseudo SU(3) model, 
Nucl. Phys. A \textbf{202}, 433 (1973). 

\bibitem{Yoshida}
K. Sugawara-Tanabe,  A. Arima, and N. Yoshida, Resurrection of the L-S coupling scheme in superdeformation,  Phys. Rev. C \textbf{51}, 1809 (1995).

\end{thebibliography}
\end{document}